\documentclass[prb,twocolumn,showpacs,preprintnumbers,amsmath,amssymb]{revtex4}

\usepackage{graphicx}
\usepackage{dcolumn}
\usepackage{bm}

\newcommand{\mapright}[1]{\smash{\mathop{\hbox to 1.0cm{\rightarrowfill}}\limits^{#1}}}

\begin{document}


\title{Macroscopic quantum dynamics of $\pi$-junctions with ferromagnetic insulators}

\author{Shiro Kawabata,$^{1,2}$ Satoshi Kashiwaya,$^3$ Yasuhiro Asano,$^4$ Yukio Tanaka,$^5$ and Alexander A. Golubov$^1$}
\affiliation{%
$^1$Faculty of Science and Technology, University of Twente, 
P.O. Box 217, 7500 AE Enschede, The Netherlands \\
$^2$Nanotechnology Research Institute (NRI), National Institute of 
Advanced Industrial Science and Technology (AIST), Tsukuba, 
Ibaraki, 305-8568, Japan \\
$^3$Nanoelectronics Research Institute (NeRI), National Institute of 
Advanced Industrial Science and Technology (AIST), Tsukuba, 
Ibaraki, 305-8568, Japan \\
$^4$Department of Applied Physics, Hokkaido University,
Sapporo, 060-8628, Japan\\
$^5$Department of Applied Physics, Nagoya University,
Nagoya, 464-8603, Japan
}%

\date{\today}

\begin{abstract}

We theoretically investigate the macroscopic quantum dynamics of a $\pi$ junction with a superconductor (S) and a multiferroic material or a ferromagnetic insulator (FI). 
By deriving the effective action from a microscopic Hamiltonian, a $\pi$-junction qubit (a S-FI-S superconducting quantum interference device ring) is proposed. 
In this qubit, a quantum two-level system is spontaneously generated and the effect of the quasiparticle dissipation is found to be very weak. 
These features make it possible to realize a quiet qubit with high coherency.
We also investigate macroscopic quantum tunneling (MQT) in current-biased S-FI-S $\pi$ junctions and show that the influence of the quasiparticle dissipation on MQT is negligibly small.
\end{abstract}

\pacs{74.50.+r, 03.67.Lx, 03.65.Yz, 74.78.Na}
\maketitle


When two superconductors are weakly coupled via a thin insulating barrier, a direct current can flow even without bias voltage.
The driving force of this supercurrent is the phase difference in the macroscopic wave function.
The supercurrent $I$ and the phase difference $\phi$ across the junction have a relation $I=I_C \sin \phi$ with $I_C>0$ being the critical current.
If the weak link consists of a thin ferromagnetic metal (FM) layer, the result can be a Josephson junction with a built-in phase difference of $\pi$.
Physically this is a consequence of the phase change of the order parameter induced in the FM by the proximity effect.~\cite{rf:Kulik,rf:Bulaevskii,rf:Buzdin} 
Superconductor (S)-FM-S Josephson junctions presenting a negative coupling or a negative $I_C$ are usually called $\pi$ junctions~\cite{rf:Golubov04,rf:BuzdinRMP} and  such behavior has been reported experimentally.~\cite{rf:Exp1,rf:Exp2,rf:Exp3,rf:Exp4,rf:Exp5}

As proposed by Bulaevskii $et$ $al.$,~\cite{rf:Bulaevskii} a superconducting ring with a $\pi$ junction [a $\pi$ superconducting quantum interference device (SQUID)] exhibits a spontaneous current without an external magnetic field and the corresponding magnetic flux is half a flux quantum $\Phi_0$ in the ground state.
Therefore it is expected that a S-FM-S $\pi$ SQUID system becomes a $quiet$ qubit that can be efficiently decoupled from the fluctuation of the external field.~\cite{rf:Ioffe,rf:Blatter,rf:Yamashita,rf:Yamashita2}
From the viewpoint of quantum dissipation, however, the structure of S-FM-S junctions is inherently identical with S-N-S junctions (where N is a normal nonmagnetic metal).
Therefore a gapless quasiparticle excitation in the FM layer is inevitable.
This feature gives a strong Ohmic dissipation~\cite{rf:Zaikin,rf:Krasnov} and the coherence time of S-FM-S qubits is bound to be very short.
In practice the current-voltage characteristic of a S-FM-S junction shows nonhysteretic and overdamped behaviors.~\cite{rf:Exp3}
On the other hand, as was shown by Tanaka and Kashiwaya,~\cite{rf:Tanaka} a $\pi$ junction can also be formed in Josephson junctions with a ferromagnetic insulator (FI).
In S-FI-S junctions, the quasiparticle excitation in the FI is expected to be very weak as in the case of S-I-S junctions~\cite{rf:Ambegaoker} (where I is a nonmagnetic insulator).

In this paper, we propose a $\pi$-junction qubit that consists of a rf SQUID ring with S-FI-S junctions, and investigate macroscopic quantum tunneling (MQT) in a single S-FI-S junction.
Unlike previous phenomenological studies for S-FM-S junction qubits,~\cite{rf:Ioffe,rf:Blatter,rf:Yamashita,rf:Yamashita2} we derive the effective action of S-FI-S junctions from a microscopic Hamiltonian in order to deal with the quasiparticle dissipation explicitly.
By using the effective action, we show that the quasiparticle dissipation in this system is considerably weaker than in S-FM-S junctions. 
This feature makes it possible to realize highly coherent quantum logic circuits.

\begin{figure}[b]
\begin{center}
\includegraphics[width=8.5cm]{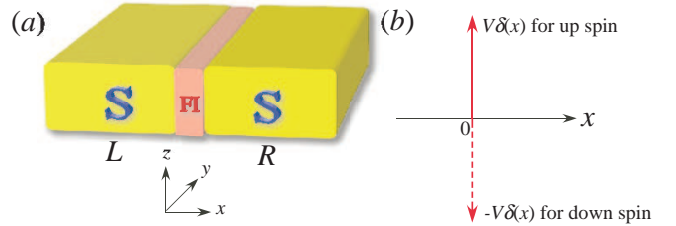}
\caption{(a) Schematic view of the superconductor-ferromagnetic insulator-superconductor (S-FI-S) Josephson junction and (b) the spin-dependent barrier potential for the FI layer.}
\end{center}
\end{figure}
First, we will calculate the effective action for S-FI-S Josephson junctions by using the functional integral method.~\cite{rf:Ambegaoker,rf:Kawabata1,rf:Kawabata2}
S-FI-S Josephson junctions consist of two superconductors (L and R) and a thin FI barrier [Fig. 1(a)].
The Hamiltonian of S-FI-S junctions is conveniently given by ${\cal H}= {\cal H}_L + {\cal H}_R + {\cal H}_T + {\cal H}_Q$, where ${\cal H}_{L(R)}$ is the Hamiltonian describing the left (right) superconductor electrodes:
$
{\cal H}_{L}
=
\sum_\sigma \int d \mbox{\boldmath $r$} 
\ 
\psi_{L \sigma}^\dagger \left( \mbox{\boldmath $r$}  \right)
\left( - \hbar^2 \nabla^2 /2 m - \mu \right)
\psi_{L \sigma}\left( \mbox{\boldmath $r$}  \right)
 -
(g_{L}/2) 
\sum_{\sigma} 
$
$
\int d \mbox{\boldmath $r$}   
\psi_{L \sigma}^\dagger \left( \mbox{\boldmath $r$}  \right)
\psi_{L -\sigma}^\dagger \left( \mbox{\boldmath $r$}  \right)
\psi_{L -\sigma} \left( \mbox{\boldmath $r$}  \right)
 \psi_{L \sigma} \left( \mbox{\boldmath $r$}  \right)
,
$
where $\psi_\sigma$ is the  electron field operator for the spin $\sigma(=\uparrow, \downarrow)$, $m$ is the electron mass, and $\mu$ is the chemical potential.
The coupling between two superconductors is due to the transfer of electrons through the FI barrier and due to the Coulomb interaction term 
$
{\cal H}_Q=({\cal Q}_L -{\cal Q}_R )^2/8 C,
$
where $C$ is the capacitance of the junction and 
$
{\cal Q}_{L(R)} = e \sum_\sigma \int  d \mbox{\boldmath $r$} \psi_{L(R) \sigma}^\dagger \left( \mbox{\boldmath $r$}  \right) \psi_{L(R) \sigma} \left( \mbox{\boldmath $r$}  \right)
$
is the operator for the charge on the superconductor L(R).
The former is described by the tunneling term
$
{\cal H}_T =
\sum_{\sigma} 
\int d \mbox{\boldmath $r$} d \mbox{\boldmath $r$}'  
[
T_{\sigma} \left( \mbox{\boldmath $r$},\mbox{\boldmath $r$}' \right) 
\psi_{L \sigma}^\dagger \left( \mbox{\boldmath $r$}  \right)
\psi_{R \sigma} \left( \mbox{\boldmath $r$}' \right)
+
\mbox{H.c.}
].
$
The FI barrier can be described by a potential,~\cite{rf:Tanaka,rf:Kashiwaya,rf:Barash} $V_\sigma (\mbox{\boldmath $r$})
=
\rho_\sigma V \delta(x)
$, where $\rho_{\uparrow} =1$ and $\rho_{\downarrow} =-1$ [see Fig. 1(b)].
In the high-barrier limit ($Z \equiv m V / \hbar^2 k_F \gg 1$), the tunneling matrix element is given by
$
 T_{\sigma} (\mbox{\boldmath $k$},\mbox{\boldmath $k$}')
=
 i \rho_\sigma k_x/(   k_F Z) \delta_{k_y,k_y'} \delta_{k_z,k_z'}
$,
where $k_F$ is the Fermi wave number.
The spin dependence of $T_{\sigma} $ is essential for the formation of $\pi$ coupling.

Examples of FIs include the $f$-electron systems Eu{\it X} ({\it X}=O, S, and Se),~\cite{rf:EuO} ferrites,~\cite{rf:Ferrite} rare-earth nitrides (e.g., GdN),~\cite{rf:Xiao,rf:GdN} insulating barriers with magnetic impurities~\cite{rf:Shiba} (e.g., amorphous FeSi alloys),~\cite{rf:Vavra} Fe-filled semiconductor carbon nanotubes,~\cite{rf:CN} and single molecular magnets (e.g., Mn${}_{12} $ derivatives)~\cite{rf:Heersche}.
Multiferroic materials,~\cite{rf:Multiferroic,rf:Khomskii} for instance, the Jahn-Teller orbital-ordered systems (e.g., Ti oxides~\cite{rf:TiO} and Mn oxides~\cite{rf:MnO,rf:Gajek}), and the spinels (e.g.,  CdCr${}_2$S${}_4$~\cite{rf:Spinels} and CoCr${}_2$O${}_4$~\cite{rf:Yamasaki}) can serve as a FI. 
 It has been recently shown theoretically that a FI can be also induced by doping in wide band-gap semiconductors such as ZnO and GaN.~\cite{rf:ZnO}
However, it is still an open question whether any of these materials posses the spin dependent potential shown in Fig.1 (b).
This problem will be addressed in a future study.

The partition function ${\cal Z}$ of the junction can be written as an imaginary time path integral over the complex Grasmmann fields,~\cite{rf:Nagaosa}
$
{\cal Z}
=
\int 
{\cal D} \bar{\psi} {\cal D} \psi 
\exp
(
  -  \int_{0}^{\hbar \beta} d \tau {\cal  L} [\tau]/\hbar
)
,
$
where $\beta=$$1/k_BT$ and the Lagrangian is given by $ {\cal L}[\tau]=\sum_\sigma\int d \mbox{\boldmath $r$} \bar{\psi}_\sigma (\mbox{\boldmath $r$},\tau) \partial_\tau  \psi_\sigma (\mbox{\boldmath $r$},\tau) + {\cal H}[\tau]$.
In order to write the partition function as a functional integral over the macroscopic variable (the phase difference $\phi$), we apply the Stratonovich-Hubbard transformation.
This introduces a complex order parameter field $\Delta(\mbox{\boldmath $r$},\tau)$.
Next the integrals over the Grassmann fields and $|\Delta| \equiv \Delta_0$ are performed by using the Gaussian integral and the saddle point approximation, respectively.
Then we obtain the partition function as 
$
Z
= 
\int 
{\cal  D} \phi (\tau) 
\exp
\left(
  -{\cal  S}_\mathrm{eff}[\phi]/\hbar
\right)
$, where the effective action ${\cal  S}_\mathrm{eff}$ is given by 
\begin{eqnarray}
{\cal  S}_\mathrm{eff}[\phi]
&= &
\int_{0}^{\hbar \beta} d \tau 
\left[
   \frac{C}{2} 
   \left(
   \frac{\hbar}{2e} 
   \frac{\partial \phi ( \tau) }{\partial \tau}
   \right)^2
   - E_J \cos \phi( \tau) 
\right]
\nonumber\\
&+ &{\cal  S}_\alpha[\phi],
\\
{\cal  S}_\alpha[\phi]
&\equiv&
-
\sum_\sigma
\int_{0}^{\hbar \beta} d \tau 
 d \tau'
  \alpha_\sigma (\tau - \tau') 
  e^{i \rho_\sigma  \frac{  \phi(\tau) - \phi (\tau') }{2}}
  \label{eqn:alpha}
.
\end{eqnarray}
Here the Josephson coupling energy $E_J= (\hbar/2e) I_C$ is given in terms of the anomalous Green's function in the left (right) superconductor ${\cal F}_{L(R)} \left( \mbox{\boldmath $k$},\omega_n \right)=\hbar \Delta_0/[(\hbar \omega_n)^2 + \xi_{\mbox{\boldmath $k$}}^2 +  \Delta_0^2]$ ($\xi_{\mbox{\boldmath $k$}}=\hbar^2 \mbox{\boldmath $k$}^2/2 m -\mu$ and $\hbar \omega_n= (2n+1) \pi /\beta$ is the fermionic Matsubara frequency):
\begin{eqnarray}
 E_J &=&
  \frac{2}{\hbar}
\int_{0}^{\hbar \beta}
d \tau 
  \sum_{\mbox{\boldmath $k$},\mbox{\boldmath $k$}'}
   T^*_{\downarrow} (\mbox{\boldmath $k$},\mbox{\boldmath $k$}')
   T_{\uparrow} (\mbox{\boldmath $k$},\mbox{\boldmath $k$}')
   \nonumber\\
   &\times&
{\cal F}_L \left( \mbox{\boldmath $k$},\tau \right)
{\cal F}_R \left( \mbox{\boldmath $k$}',-\tau \right)
 \approx  - \frac{\Delta_0 R_Q}{4 \pi R_N} < 0.
\end{eqnarray}
In this equation, $R_Q= h/4 e^2$ is the resistance quantum, and $R_N$ is the normal state resistance of the junction.
As expected, $E_J$ becomes negative.
The formation of the $\pi$ junction can be attributed to the spin-discriminating scattering processes in the spin-dependent potential $V_\sigma (\mbox{\boldmath $r$})$. 
Therefore S-FI-S junctions can serve as $\pi$ junctions similar to S-FM-S junctions.
${\cal S}_\alpha$ is the dissipation action and describes the tunneling of quasiparticles which is the origin of the quasiparticle dissipation. 
In Eq. (\ref{eqn:alpha}), the dissipation kernel $\alpha_\sigma(\tau)$ is given by
\begin{eqnarray}
  \alpha_{\sigma}  (\tau )  =
  -\frac{2}{\hbar}
  \sum_{\mbox{\boldmath $k$},\mbox{\boldmath $k$}'}
  \left|
   T_{\sigma} (\mbox{\boldmath $k$},\mbox{\boldmath $k$}')
  \right|^2
 {\cal G}_L \left( \mbox{\boldmath $k$},\tau \right)
{\cal G}_R \left( \mbox{\boldmath $k$}',-\tau \right)
 ,
\end{eqnarray}
where ${\cal G}_{L(R)}\left( \mbox{\boldmath $k$},\omega_n \right)=-\hbar ( i \hbar \omega_n + \xi_{\mbox{\boldmath $k$}} )/[(\hbar \omega_n)^2 + \xi_{\mbox{\boldmath $k$}}^2 +  \Delta_0^2]$ is the diagonal component of the Nambu Green's function.
In the high-barrier limit ($Z \gg 1$), we obtain
\begin{eqnarray}
  \alpha_{\sigma}  (\tau ) =
\frac{\Delta^2}{ 4 \pi^2 e^2 R_N} K_1 \left(  \frac{\Delta |\tau|}{\hbar} \right)^2
,
\label{eqn:alpha2}
\end{eqnarray}
where $K_1$ is the modified Bessel function.
For $ |\tau| \gg \hbar / \Delta_0$ the dissipation kernel decays exponentially as a function of the imaginary time $\tau$, i.e., $\alpha_{\sigma}$$ (\tau ) \sim \exp \left(-  2 \Delta_0 |\tau|  / \hbar \right)$.
If the phase varies only slowly with the time scale given by $\hbar /\Delta_0$, we can expand $\phi(\tau) - \phi (\tau')$ in Eq. (\ref{eqn:alpha}) about $\tau=\tau'$.
This gives 
$
{\cal S}_\alpha[\phi]
\approx
(\delta C/2)
\int_{0}^{\hbar \beta} d \tau 
   \left[
   (\hbar/2e)
   \partial \phi ( \tau) /\partial \tau
   \right]^2
$.
Hence the dissipation action ${\cal S}_\alpha$ acts as a kinetic term so that the effect of the quasiparticles results in an increase of the capacitance, $C \to C + \delta C$.
This indicates that the quasiparticle dissipation in S-FI-S junctions is qualitatively weaker than that in S-FM-S junctions in which the strong Ohmic dissipation appears.~\cite{rf:Zaikin,rf:Krasnov}
At zero temperature, the capacitance increment $\delta C$ can be easily calculated using Eq. (\ref{eqn:alpha2}) and we can obtain
\begin{eqnarray}
\delta C = \frac{3}{32 \pi} \frac{e^2 R_Q}{\Delta_0 R_N}
\label{eqn:deltaC}
.
\end{eqnarray}
As will be shown in later, $\delta C/C \ll 1$. 
Therefore the effect of the quasiparticle dissipation on the quantum dynamics of S-FI-S junctions is very small.

By using the above result, we propose a different type of flux qubit.~\cite{rf:Mooij}
In Fig. 2(a), we show the schematic of the $\pi$ SQUID qubit.
In this proposal, the qubit consists of the superconducting rf SQUID loop (the inductance $L_\mathrm{loop}$) with one S-FI-S junction.
The effective Hamiltonian which describes this qubit is given by
\begin{eqnarray}
{\cal H}_\mathrm{eff} = \frac{C_\mathrm{ren}}{2}
\left(
\frac{\hbar}{2 e} 
\frac{\partial \phi}{\partial t}
\right)^2
+ U (\phi)
.
\end{eqnarray}
Here $C_\mathrm{ren} \equiv C + \delta C$ and $U (\phi)$ is the potential energy:
\begin{eqnarray}
U (\phi)
=
 - 
 E_J  \cos \phi
+
\frac{1}{2 L_\mathrm{loop}}
\left( \frac{\Phi_0}{2 \pi} \right)^2
\left( \phi - 2 \pi \frac{\Phi_\mathrm{ex}}{\Phi_0} \right)^2
,
\end{eqnarray}
where $\Phi_\mathrm{ex}$ is the external magnetic flux.
$U(\phi)$ exhibits two minima at $\phi=\pm \pi/2$ without a external magnetic flux $\Phi_\mathrm{ex}$ [see Fig. 2(b)] and has two identical wells with equal energy
levels when tunneling between the wells is neglected. 
These levels correspond to clockwise and counterclockwise persistent currents circulating in the loop (the half flux states).
Let us consider the lowest (doubly degenerate) energy levels ($|\uparrow \rangle$ and $| \downarrow \rangle$). 
When the tunneling between two wells is switched on, the levels split, and a two-level system ($| 0 \rangle = (| \uparrow \rangle +| \downarrow \rangle)/\sqrt{2}$ and $| 1 \rangle = (| \uparrow \rangle - | \downarrow \rangle)/\sqrt{2}$) is formed with the level spacing $\epsilon$.
These states are used as a computational basis for the qubit.~\cite{rf:comment}
\begin{figure}[t]
\begin{center}
\includegraphics[width=8.5cm]{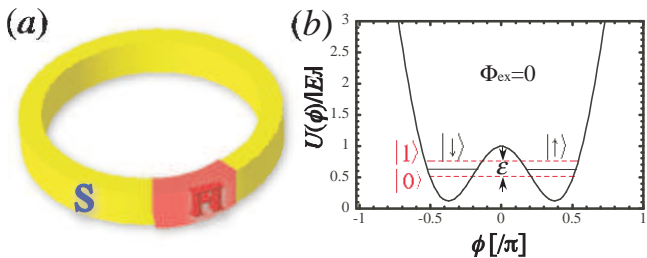}
\caption{(a) Schematics of the $\pi$ SQUID flux qubit with a single S-FI-S junction. (b) The potential energy $U(\phi)$ vs the phase difference $\phi$ without the external magnetic flux $\Phi_\mathrm{ex}$.}
\end{center}
\end{figure}
In order to prevent thermalization of the quantum state composed from the two low-lying energy levels ($| 0 \rangle$ and $| 1 \rangle$), we require $k_B T \ll  \epsilon$.

As in the case of the qubit using the S-I-S flux qubit,~\cite{rf:Mooij} the one-qubit operation (the Rabi oscillation) can be realized by irradiating the qubit with microwaves of the frequency $\hbar \omega = \epsilon$.
Moreover, two-qubit gate  (e.g., the controlled-NOT gate) can be performed by using the inducting coupling between two adjacent qubits.

Next, we will develop a theory of MQT in single S-FI-S junctions [Fig. 1(a)].
MQT is an important first step toward the experimental realization of Josephson junction qubits and is used in a final measurement process for a phase-type qubit.~\cite{rf:Han,rf:Martinis,rf:Kawabata3}
In order to observe MQT, an external bias current $I_\mathrm{ext}$ which is close to $I_C$ is applied to the junction.
This leads to an additional term $-(\hbar/2e) \int_0^{\hbar \beta} d \tau I_\mathrm{ext} \phi (\tau)$ in the effective action (1).~\cite{rf:Ambegaoker}
The resultant action 
\begin{eqnarray}
{\cal S}_\mathrm{eff}[\phi]
&= &
\int_{0}^{\hbar \beta} d \tau 
\left[
   \frac{C_\mathrm{ren}}{2} 
   \left(
   \frac{\hbar}{2e} 
   \frac{\partial \phi ( \tau) }{\partial \tau}
   \right)^2
   -U(\phi)\right]
\end{eqnarray}
describes the quantum dynamics of a fictive particle (the macroscopic phase difference $\phi$) with mass $M=C_\mathrm{ren}(\hbar/2e)^2$ moving in the tilted washboard potential $U(\phi)=  - E_J  \left[ \cos \phi( \tau)  - \eta \phi(\tau) \right]$, where $\eta \equiv I_\mathrm{ext}/ |I_C |$.
The MQT escape rate from this metastable potential at zero temperature is given by
$
\Gamma
=
\lim_{\beta \to \infty} (2/\beta) \mbox{ Im}\ln {\cal Z}.
$
~\cite{rf:MQT1}
By using the semiclassical (instanton) method,~\cite{rf:Caldeira} the MQT rate is approximated as 
\begin{eqnarray}
\Gamma(\eta)
=
\frac{\omega_p(\eta)}{2 \pi} \sqrt{120 \pi B(\eta) }\  e^{ -B(\eta)},
  \label{eqn:MQT}
\end{eqnarray}
where $\omega_p (\eta)= \sqrt{\hbar I_C / 2 e M}(1-\eta^2)^{1/4}$ is the Josephson plasma frequency and $B(\eta)= {\cal S}_{\mathrm{eff}}[\phi_B]/\hbar$ is the bounce exponent, that is, the value of the action ${\cal S}_{\mathrm{eff}}$ evaluated along the bounce trajectory $\phi_B(\tau)$.
The analytic expression for the exponent $B$ is given by
\begin{eqnarray}
B(\eta)=\frac{12}{5 e}
  \sqrt{ \frac{\hbar}{2 e} I_C C_\mathrm{ren}}
  \left(
  1 -
 \eta^2
\right)^{\frac{5}{4}}
.
\end{eqnarray}

In MQT experiments, the switching current distribution $P(\eta)$ is measured. 
$P(\eta)$ is related to the MQT rate $\Gamma(\eta)$ as 
\begin{eqnarray}
P(\eta)=\frac{1}{v}
\Gamma(\eta) \exp
\left(
- \frac{1}{v}
\int_0^{\eta} \Gamma(\eta') d \eta'
\right)
,
\end{eqnarray}
where $v \equiv \left| d \eta / d t \right| $ is the sweep rate of the external bias current.
The average value of the switching current is expressed by $\langle \eta \rangle \equiv \int_0^1 d \eta' P(\eta') \eta'$.
At high temperature regime, the thermally activated decay dominates the escape process.
Then the escape rate is given by the Kramers formula~\cite{rf:MQT1}
$
\Gamma=(\omega_p/2 \pi )\exp \left( - U_0/ k_B T\right),
$
where $U_0$ is the barrier height.
Below the crossover temperature $T^*$, the escape process is dominated by MQT and the escape rate is given by Eq. (\ref{eqn:MQT}). 
The crossover temperature $T^*$ is defined by~\cite{rf:Kato}
\begin{eqnarray}
T^*
=
\frac{ 5 \hbar \omega_p (\eta=\langle \eta \rangle)}{36 k_B}
  \label{eqn:Tco}
.
\end{eqnarray}
As was shown by Caldeira and Leggett, in the presence of a dissipation, $T^*$ is suppressed.~\cite{rf:Caldeira}

In order to see explicitly the effect of the quasiparticle dissipation on MQT, we numerically estimate $T^*$.
Currently no experimental data are available for S-FI-S junctions.
Therefore we estimate $T^*$ by using the parameters for a high-quality Nb/Al${}_2$O${}_3$/Nb junction~\cite{rf:Ustinov} ($\Delta_0=1.30$meV,  $C=1.61$pF, $ |I_C | = 320 \mu$A,  $R_N=\Delta_0/4 e  |I_C |$, $ v  |I_C | = 0.245$A/s).
By substituting these data into Eq. (\ref{eqn:deltaC}) we obtain $\delta C/C = 0.0145 \ll 1$.
Then from Eq. (\ref{eqn:Tco}) we get the crossover temperature $T^* = 245 $mK for the dissipationless case ($C_\mathrm{ren}= C$) and $T^* = 244 $mK for the dissipation case ($C_\mathrm{ren}= C + \delta C$).
We find that, due to the existence of the quasiparticle dissipation, $T^*$ is reduced, but this reduction is negligibly small.
This strongly indicates the high potentiality for the S-FI-S junctions as a phase-type qubit.~\cite{rf:Han,rf:Martinis,rf:Kawabata3}

To summarize, we have theoretically proposed a $\pi$-junction quiet qubit which consists of a superconducting ring with the FI (the S-FI-S $\pi$ SQUID qubit).
Moreover, we have investigated the effect of the quasiparticle dissipation on the quantum dynamics and MQT using the parameter set for a high-quality Nb junction with Al${}_2$O${}_3$ barrier, and showed that this effect is considerably smaller compared with S-FM-S junction cases.
This feature and the quietness of this system make it possible to realize a quiet qubit  with long coherence time.

Finally we would like to comment on the possibility of a quiet qubit using a S-I-FM-S junction.~\cite{rf:Golubov04}
Recently Weides $et$ $al.$~\cite{rf:Weides0,rf:Weides1,rf:Weides2} and Born $et$ $al.$~\cite{rf:Born} have fabricated high-quality S-I-FM-S junctions, i.e., Nb/Al${}_2$O${}_3$/Ni${}_{0.6}$Cu${}_{0.4}$/Nb and Nb/Al/Al${}_2$O${}_3$/Ni${}_{3}$Al/Nb, respectively.
They have clearly observed the 0-$\pi$ transitions by changing the thickness of the FM layer.
In these systems, the quasiparticle tunneling is inhibited due to the existence of the insulating barrier I (Al${}_2$O${}_3$).
Therefore, as in the case of S-FI-S junctions, low quasiparticle dissipation and quietness are also expected in S-I-FM-S junctions.
The theory of the qubit and MQT in such systems will be the subject of future studies.

We  would like to thank J. Aarts, H. Akinaga, C. Bell, H. Ito, T. Kato, P. J. Kelly, J. Nitta, S. Takahashi, T. Yamashita and M. Weides for useful discussions.
One of the authors (Kawabata) would like to express deep gratitude to S. Abe and  H. Yokoyama for their encouragement.
This work was partly supported by NEDO under the Nanotechnology Program, a Grant-in-Aid for Scientific Research from the Ministry of Education, Science, Sports and Culture of Japan (grant No. 17710081), and the Nano-NED Program under project No. TCS.7029.


\begin{thebibliography}{99}
%
%
%
%
\bibitem{rf:Kulik}
I. O. Kulik, 
Sov. Phys. JETP {\bf 22}, 841 (1966).
%
%
\bibitem{rf:Bulaevskii}
L. N. Bulaevskii, V. V. Kuzii, and A. A. Sobyanin, 
JETP Lett. {\bf 25}, 291 (1977).
%
%
\bibitem{rf:Buzdin}
A. I. Buzdin, L. N. Bulaevskii, and S. V. Panyukov, 
JETP Lett. {\bf 35}, 179 (1982).
%
%
\bibitem{rf:Golubov04}
A. A. Golubov, M. Yu. Kupriyanov, and E. Il'ichev, 
Rev. Mod. Phys. {\bf 76}, 411 (2004).
%
%
\bibitem{rf:BuzdinRMP}
A. I. Buzdin,
Rev. Mod. Phys. {\bf 77}, 935 (2005).
%
%
\bibitem{rf:Exp1}
V. V. Ryazanov {\it et al.}, 
Phys. Rev. Lett. {\bf 86}, 2427 (2001).
%
%
\bibitem{rf:Exp2}
T. Kontos {\it et al.}, 
Phys. Rev. Lett. {\bf 89}, 137007 (2002).
%
%
\bibitem{rf:Exp3}
S. M. Frolov {\it et al.}, 
Phys. Rev. B {\bf 70}, 144505 (2004).
%
%
\bibitem{rf:Exp4}
H. Sellier {\it et al.}, 
Phys. Rev. Lett. {\bf 92}, 257005 (2004).
%
%
\bibitem{rf:Exp5}
J. W. A. Robinson {\it et al.}, 
Phys. Rev. Lett. {\bf 97}, 177003 (2006).
%
%
\bibitem{rf:Ioffe}
L. B. Ioffe {\it et al.}, 
Nature  (London) {\bf 398}, 679 (1999).
%
%
\bibitem{rf:Blatter}
G. Blatter, V. B. Geshkenbein, and L. B. Ioffe,
Phys. Rev. B {\bf 63}, 174511 (2001).
%
%
\bibitem{rf:Yamashita}
T. Yamashita {\it et al.}, 
Phys. Rev. Lett. {\bf 95}, 097001 (2005).
%
%
\bibitem{rf:Yamashita2}
T. Yamashita, S. Takahashi, and S. Maekawa, 
Appl. Phys. Lett. {\bf 88}, 132501 (2006).
%
%
\bibitem{rf:Zaikin}
A. D. Zaikin and S. V. Panyukov, 
Sov. Phys. JETP {\bf 62}, 137 (1985).
%
%
\bibitem{rf:Krasnov}
V. M. Krasnov {\it et al.}, 
Phys. Rev. Lett. {\bf 95}, 157002 (2005).
%
%
\bibitem{rf:Tanaka}
Y. Tanaka and S. Kashiwaya,
Physica C {\bf 274}, 357 (1997).
%
%
\bibitem{rf:Ambegaoker}
U. Eckern, G. Sch\"on, and V. Ambegaokar, 
Phys. Rev. B  {\bf 30}, 6419 (1984).
%
%
\bibitem{rf:Kawabata1}
S. Kawabata {\it et al.}, 
Phys. Rev. B  {\bf 70}, 132505 (2004).
%
%
\bibitem{rf:Kawabata2}
S. Kawabata {\it et al.}, 
Phys. Rev. B  {\bf 72}, 052506 (2005).
%
%
\bibitem{rf:Kashiwaya}
S. Kashiwaya {\it et al.}, 
Phys. Rev. B  {\bf 60}, 3572 (1999).
%
%
%
\bibitem{rf:Barash}
Yu. S. Barash and  I. V. Bobkova, 
Phys. Rev. B  {\bf 65}, 144502 (2002).
%
%
\bibitem{rf:EuO}
R. Meservey and P. M. Tedrow,
Phys. Rep. {\bf 238}, 173 (1994).
%
%
\bibitem{rf:Ferrite}
A. Goldman, {\it Modern Ferrite Technology}  (Springer-Verlag, New York, 2006). 
%
%
\bibitem{rf:Xiao}
J. Q. Xiao and C. L. Chien,
Phys. Rev. Lett. {\bf 76}, 1727 (1996).
%
%
\bibitem{rf:GdN}
S. Granville {\it et al.}, 
Phys. Rev. B  {\bf 73}, 235335 (2006).
%
%
\bibitem{rf:Shiba}
H. Shiba and T. Soda,
Prog. Theor. Phys. {\bf 41}, 25 (1969).
%
%
\bibitem{rf:Vavra}
O. V\'avra {\it et al.}, 
Phys. Rev. B  {\bf 74}, 020502(R) (2006).
%
%
\bibitem{rf:CN}
Y. F. Li {\it et al.}, 
Appl. Phys. Lett. {\bf 89}, 083117 (2006).
%
%
\bibitem{rf:Heersche}
H. B. Heersche {\it et al.}, 
Phys. Rev. Lett. {\bf 96}, 206801 (2006).
%
%
\bibitem{rf:Multiferroic}
W. Eerenstein, N. D. Mathur, and J. F. Scott,
Nature (London) {\bf 442}, 759 (2006).
%
%
\bibitem{rf:Khomskii}
D. I. Khomskii, 
J. Magn. Magn. Mater. {\bf 306}, 1 (2006).
%
%
\bibitem{rf:TiO}
M. Mochizuki and M. Imada,
New J. Phys. {\bf 6}, 154 (2004).
%
%
\bibitem{rf:MnO}
Y. Tokura and N. Nagaosa,
Science {\bf 288}, 462 (2000).
%
%
\bibitem{rf:Gajek}
M. Gajek {\it et al.}, 
cond-mat/0606444.
%
%
\bibitem{rf:Spinels}
P. K. Baltzer, H. W. Lehmann, and M. Robbins,
Phys. Rev. Lett. {\bf 15}, 493 (1965).
%
%
\bibitem{rf:Yamasaki}
Y. Yamasaki {\it et al.}, 
Phys. Rev. Lett. {\bf 96}, 207204 (2006).
%
%
\bibitem{rf:ZnO}
T. Dietl {\it et al.}, 
Science {\bf 287}, 1019 (2000).
%
%
\bibitem{rf:Nagaosa}
N. Nagaosa, {\it Quantum Field Theory in Condensed Matter Physics}  (Springer-Verlag, Berlin, 1999). 
%
%
\bibitem{rf:Mooij}
J. E. Mooij {\it et al.}, 
Science {\bf 285}, 1036 (1999). 
%
%
\bibitem{rf:comment}
 From a practical point of view, one drawback of the flux qubit with a single $\pi$ junction described in this paper concerns the large inductance $L_\mathrm{loop}$, the energy of which must be comparable to $|E_J|$ to form the required double-well potential profile. This implies a large size of the qubit loop, which makes the qubit vulnerable to decoherence by magnetic fluctuations of the environment. 
One way to overcome this difficulty is usage of one $\pi$ junction and two S-I-S 0 junctions connected in series in a superconducting ring.~\cite{rf:Yamashita2,rf:Mooij}
 The inductive energy of the ring is chosen to be much smaller than $|E_J|$ of the junctions.
%
%
\bibitem{rf:Han}
Y. Yu {\it et al.}, 
Science {\bf 296}, 889 (2002). 
%
%
\bibitem{rf:Martinis}
J. M. Martinis {\it et al.}, 
Phys. Rev. Lett. {\bf 89}, 117901 (2002).
%
%
\bibitem{rf:Kawabata3}
S. Kawabata {\it et al.}, 
Physica E (Amsterdam) {\bf 29}, 669 (2005).
%
%
\bibitem{rf:MQT1}
U. Weiss, {\it Quantum Dissipative Systems}  (World Scientific, Singapore, 1999).
%
%
\bibitem{rf:Caldeira}
A. O. Caldeira and A. J. Leggett, 
Ann. Phys. (N.Y.) {\bf 149}, 374 (1983).
%
%
\bibitem{rf:Kato}
T. Kato and M. Imada
J. Phys. Soc. Jpn. {\bf 65}, 2963 (1996).
%
%
\bibitem{rf:Ustinov}
A. Wallraff {\it et al.}, 
Rev. Sci. Instrum. {\bf 74}, 3740 (2003).
%
%
\bibitem{rf:Weides0}
M. Weides, K. Tillmann, and H. Kohlstedt,
Physica C {\bf 437-438}, 349 (2006).
%
%
\bibitem{rf:Weides1}
M. Weides {\it et al.}, 
Appl. Phys. Lett. {\bf 89}, 122511 (2006).
%
%
\bibitem{rf:Weides2}
M. Weides {\it et al.}, 
cond-mat/0605656.
%
%
\bibitem{rf:Born}
F. Born {\it et al.}, 
Phys. Rev.B {\bf 74}, 140501(R) (2006).
%
\end{thebibliography}
\end{document}